\documentclass[fleqn,usenatbib]{mnras}

\usepackage{newtxtext,newtxmath}
\usepackage[T1]{fontenc}

\usepackage{graphicx}	% Including figure files
\usepackage{amsmath}	% Advanced maths commands
\usepackage{newtxtext,newtxmath}

\newcommand{\jtto}{V496~UMa}
\newcommand{\xmm}{\textit{XMM-Newton}}
\newcommand{\tess}{\textit{TESS}}

\title[X-ray observations of V 496 UMa]{Transient two pole accretion in the polar V496 UMa}

\author[M. R. Kennedy et al.]{
M. R. Kennedy,$^{1,2}$\thanks{E-mail: markkennedy@ucc.ie}\thanks{These authors contributed equally to this work.}
C. Littlefield,$^{3, 4}$\footnote[2]{}
and P. M. Garnavich$^{3}$
\\
$^{1}$Department of Physics, University College Cork, Cork, Ireland\\
$^{2}$Jodrell Bank Centre for Astrophysics, Department of Physics and Astronomy, The University of Manchester, M19 9PL, UK\\
$^{3}$Department of Physics, University of Notre Dame, Notre Dame, IN 46556 USA\\
$^{4}$Bay Area Environmental Research Institute, Moffett Field, CA 94035 USA
}

\date{Accepted XXX. Received YYY; in original form ZZZ}

\pubyear{2021}

\begin{document}
\label{firstpage}
\pagerange{\pageref{firstpage}--\pageref{lastpage}}
\maketitle

\begin{abstract}

We report \xmm\ and \tess\ observations of \jtto, an AM Herculis-type cataclysmic variable. The \xmm\ observation reveals that at times, two poles on the white dwarf accrete simultaneously, but accretion onto the secondary magnetic pole is erratic and can nearly cease in less than one binary orbit (1.5~h). Modelling of the X-ray spectrum during the primary maximum reveals no change in the accretion structures onto the primary pole when accretion onto the secondary pole is disrupted, suggesting that the disruption of accretion onto the secondary pole may be caused by mass-transfer variations from the donor star. The \tess\ observation, which spanned eight weeks at a two-minute cadence, shows a stable, double-humped orbital modulation due to cyclotron emission from the post-shock region, while the observed times of maximum light show a slow systematic drift that does not correlate with the system's overall brightness. 

\end{abstract}


%%%%%%%%%%%%%%%%%%%%%%%%%%%%%%%%%%%%%%%%%%%%%%%%%%

%%%%%%%%%%%%%%%%% BODY OF PAPER %%%%%%%%%%%%%%%%%%

\section{Introduction}
Cataclysmic variables (CV) are compact binary systems with white dwarf (WD) primaries that are accreting material from a nearby companion star which fills its Roche lobe. When a new CV is discovered, it is important that a clear understanding the accretion structures within the binary is established, as this gives us information regarding the central WD. Indeed, the path material from the companion takes as it flows through the inner Lagrange point (L1) towards the WD is dictated by the magnetic field of the WD. In systems where the WDs surface magnetic field is large (>5 MG), material flows first as a ballistic stream towards the WD until the point at which the magnetic pressure exerted on the material by the WDs magnetic field overcomes the ram pressure inside the stream. From this point onwards, material couples to the WDs magnetic field and flows towards the WDs magnetic poles. Systems in which the magnetic field is strong enough to produce such an accretion structure are called AM Her stars (after the archetypal system) or polars due to the high percentage of polarsied optical light which they produce \citep{1977ApJ...212L.125T}. They are formed as, soon after the binaries formation, the rotational period of the WD ($P_{\rm{s}}$) synchronises to the orbital period ($P_{\rm{O}}$) of the system and becomes tidally locked due to the interaction between the WDs magnetic field and the secondary stars magnetic field.

Due to tidal locking, there is a preferential magnetic pole on the WD for material to flow to - the one which is aligned best with material in the ballistic stream (for the remainder of this paper, we shall refer to this as the primary pole). During the early years of polar study, accretion was thought to only occur onto a single pole of the white dwarf. Such accretion onto a single pole leads to large amplitude variations at optical and X-ray wavelengths as the accreting pole rotates in and out of our field of view. However, after the discovery of polars which underwent changes in the sign of the circular polarisation (e.g. VV Pup: \citealt{1979ApJ...229..652L}), polars which had two optical maxima and minima per orbit (e.g. EF Eri; \citealt{1980MNRAS.192..689W}),  and the variable light curve of the archetype of polars AM Her \citep{1985A&A...148L..14H}, it was quickly realised that a second pole might accrete within these systems. This idea of 2 pole accretion was soldified by spectroscopic observations of VV Pup, in which 2 distinct sets of cyclotron features (corresponding to magnetic both poles of the WD) were observed \citep{1989ApJ...342L..35W}. Since then, secondary pole accretion has become a common feature of many polars.

While accretion onto this secondary pole can be constant and uninterrupted (e.g. \citealt{1999A&A...343..157R}, \citealt{2001A&A...374..189S}, and \citealt{2002A&A...392..505S}), it can also be transient, and manifests either as a change in the optical and X-ray light curve (as in AM Her), or as a change in the sign of the circular polarisation of light coming from a polar (e.g. as in VV Pup and QQ Vul; \citealt{2000MNRAS.313..533S}). In such cases, the accretion stream can be thought of as a probe of the WDs magnetic field, as it couples on to different fields lines at different times, helping us to build a picture of the WDs magnetic field structure.

For systems with a transient behaviour, there are two possible explanations. The first is that the white dwarf is spinning with a period slightly longer or shorter than the orbital period. There are a handful of polar systems for which this true, and $P_{\rm{s}}$ is $\sim2\%$ smaller or larger than $P_{\rm{O}}$. These systems are thought to have been knocked out of synchronicity by a nova eruption on the WD, an idea developed after observations of V1500 Cyg after a nova outburst in 1975 \citep{1988ApJ...332..282S}. They are expected to synchronise after enough time has passed,and if this asynchronicity is the case of the transient two pole accretion, then the phenomenon should occur over a periodic timescale (e.g. as in clearly in \textit{TESS} observations of the asynchronous polar CD Ind; \citealt{hakala19} and \citealt{littlefield19}).

However, there are clear cases where a polar switches between one and two pole accretion, and is not asynchronous. One need look no further than the archetype of polars, AM Her, to see a firm example. AM Her has been observed in both one-pole and two-pole configurations, but the timescale for switching between configurations is months-years. For short epochs of observations ($\sim$ months), the accretion geometry seems stable (see \citealt{2020A&A...642A.134S} for a thorough review on the variability seen in AM Her).

The alternate model is that the transient behaviour is caused by a change in the mass transfer rate from the binary. When the transfer rate is high, the ram pressure within the accretion structures is high enough such that the penetration depth the ballistic stream achieves into the WDs magnetic field is deep enough for material to reach field lines connected to the second pole. If the mass transfer rate drops, the penetration depth decreases, leading to cessation of accretion onto the secondary pole. This variable accretion model led researchers to investigate whether the X-ray emission from the secondary pole is not described by the typical shock model (\citealt{1979MNRAS.188..653K};\citealt{1979ApJ...234L.117L}), but instead may be due to ``blobby'' accretion (\citealt{1982A&A...114L...4K}; \citealt{1988A&A...193..113F}), where individual blobs of accreting material penetrate below the WD photosphere, manifesting as thermal radiation. Such a model has been applied to explain the different accretion regimes within AM Her, and predicts significantly different X-ray spectra from the primary and secondary poles. (\citealt{1988MNRAS.235..433H};\citealt{2020A&A...642A.134S}).

The cause of the variation in the mass transfer rate have been explained by stellar spots on the secondary star causing a temporary change in the accretion rate \citep{1994ApJ...427..956L}, but the time scale for switching between one and two pole accretion is often on a timescale of weeks to years.

Differentiating whether two-pole accretion is occurring due to asynchronicity or a variable mass transfer rate cases requires long term monitoring to identify any periodicity in the transitions between single and two-pole accretion. Finally, determining whether a polar is undergoing ``blobby'' accretion onto the secondary pole requires X-ray spectra of both the primary and secondary poles.

This paper focuses on the polar \jtto, and on answering questions surrounding the accretion geometry and its stability. This system has been the subject of two dedicated studies, both of which reported time-series photometry and optical spectroscopy. \citet{littlefield15} measured a 91-minute orbital period and showed that a typical orbital light curve contains two photometric maxima, one of which peaks at $V\sim16.5$ and the other at $V\sim17$. A single, low-resolution spectrum showed the H, He~I, and He~II emission lines which are characteristic of a polar accreting at a high accretion rate, along with a non-thermal continuum. \citet{littlefield18} followed up with time-series spectroscopy showing that \jtto's emission-line spectrum transitions into an absorption spectrum for several minutes during each orbit when the accretion curtain eclipses the cyclotron-emitting region. They also established that the non-thermal continuum in the optical spectrum is caused by smearing of the harmonics of \jtto's cyclotron spectrum. \jtto's parallax from Gaia EDR3 \citep{gaia, gaiaEDR3} yields a distance of $760\pm30$~pc using the geometric algorithm from \citet{BJ21}.

\jtto's most distinguishing property is the intermittent nature of the secondary maximum in its optical light curve. \citet{littlefield18} found that four of the 45 secondary maxima in their photometry were either extraordinarily weak or completely absent, with no apparent impact on the rest of the orbital light curve \citep[Fig.~3 in][]{littlefield18}. When absent, \jtto\ could be as much as 2.5~mag fainter during the expected secondary maxima than its normal brightness during this part of the orbit. Even more surprisingly, a failed secondary maximum in one orbit could be followed by a normal secondary maximum in the very next orbit, establishing that the mechanism responsible for the failed maxima operates on timescales of less than one orbit. \citet{littlefield18} speculated that the missing maxima might arise from intermittent accretion onto the secondary magnetic pole but lacked the observational data to test this proposal.

Motivated by the question of what is causing the missing secondary maximum, and whether the X-ray spectrum and light curve vary in a similar manner, we obtained X-ray data of \jtto\ using the \textit{XMM-Newton} X-ray telescope. During preparation of these data (present in Section 3), \jtto\ was also observed by the Transiting Exoplanet Survey Satellite (\textit{TESS}; \citealt{tess}), allowing for a unique opportunity to probe the long term nature of the variability of the secondary maximum. These data are discussed in Section 4.

\section{Observations}

\begin{table*}
	\centering
	\caption{Details of the various observations of \jtto.}
	\label{tab:observing_log}
	\begin{tabular}{lccc} % four columns, alignment for each
		\hline\hline
		Facility & Start Time & End Time & Cadence\\
		\hline
        \textit{XMM-Newton} & 2017-12-03 08:35:31 & 2017-12-03 16:38:51 & 100s (X-ray), 10s (Optical)\\
        SLKT & 2019-08-29 01:53:49& 2019-08-29 03:46:07&33s\\
        SLKT & 2019-09-06 02:14:52 & 2019-09-06 03:44:54 & 33s\\
        SLKT &2019-09-18 01:12:35 & 2019-09-18 02:53:00 & 33s \\
        SLKT & 2017-12-03 08:14:49 & 2017-12-03 11:59:14 & 33s\\
        AAVSO & 2017-12-02 08:18:46 & 2017-12-07 12:32:30 & \\
        \textit{TESS} & 2019-08-15 & 2019-10-06 & 120 s\\
		\hline\hline
	\end{tabular}
\end{table*}

\subsection{XMM-Newton}
\jtto\ was observed by \xmm\ for 29 ks starting 2017-12-03 08:35:31 (UTC). The European Photon Imaging Camera (EPIC) -pn \citep{epic_pn}, -MOS1, and MOS2 \citep{epic_mos} instruments were all operated in full frame mode with a thin filter inserted. The Reflection Grating Spectrographs (RGS1 and RGS2; \citealt{xmm_rgs}) were both operated in spectroscopy HER+SES mode. The Optical Monitor (OM; \citealt{xmm_om}) was operated in fast imaging mode with a white filter inserted. Due to the OMs observing mode, there are brief gaps in coverage every $\sim26$ min. Initial inspection of the RGS data suggested no appreciable signal was detected, and these data will not be discussed further.

All data were reduced using tasks in \texttt{SAS} v16.1.0. All data were corrected to the solar system barycentre using \texttt{Barycen}. A background light curve was inspected to look for periods of high background which may have affected the data, but none were found. All extracted spectra and light curves are available through an online repository.

\subsection{TESS}

\textit{TESS} observed \jtto\ in two consecutive sectors at a two-minute cadence. Observations began in Sector 14 on 2019 Aug. 15 and continued until the end of Sector 15 on 2019 Oct. 6. The TESS data are nearly uninterrupted, except for three downlink gaps. Both \tess\ light curves were extracted with {\tt lightkurve} \citep{lightkurve}. After experimenting with different extraction apertures, we decided to use the pipeline apertures. Due to \tess's 24-arcsec pixels, the \tess\ observations of \jtto\ are blended. In spite of this blending, \jtto\ and its photometric variability were both readily apparent in visual inspection of the \tess\ images.

\subsection{Ground-Based Optical Photometry}
Additional observations of \jtto\ were carried out by members of the American Association of Variable Star Observers (AAVSO) in the days leading up to, during, and after the \xmm\ observations. These data were used to identify correlations between the X-ray behaviour and optical behaviour of the system.

Finally, we used the 80-cm Sarah L. Krizmanich Telescope (SLKT) at the University of Notre Dame to obtain time-series photometry of \jtto\ during the first part of the \xmm\ observation as well as three light curves while \tess\ observations were underway. Table~\ref{tab:observing_log} summarizes these observations. The observations consisted of 30-second unfiltered exposures with approximately 3~s of overhead between images. Data were debiased and flatfielded in the usual fashion, and differential aperture photometry used to extract the flux of \jtto.

\section{X-ray}
\begin{figure*}
    \centering
    \includegraphics[width=0.8\textwidth]{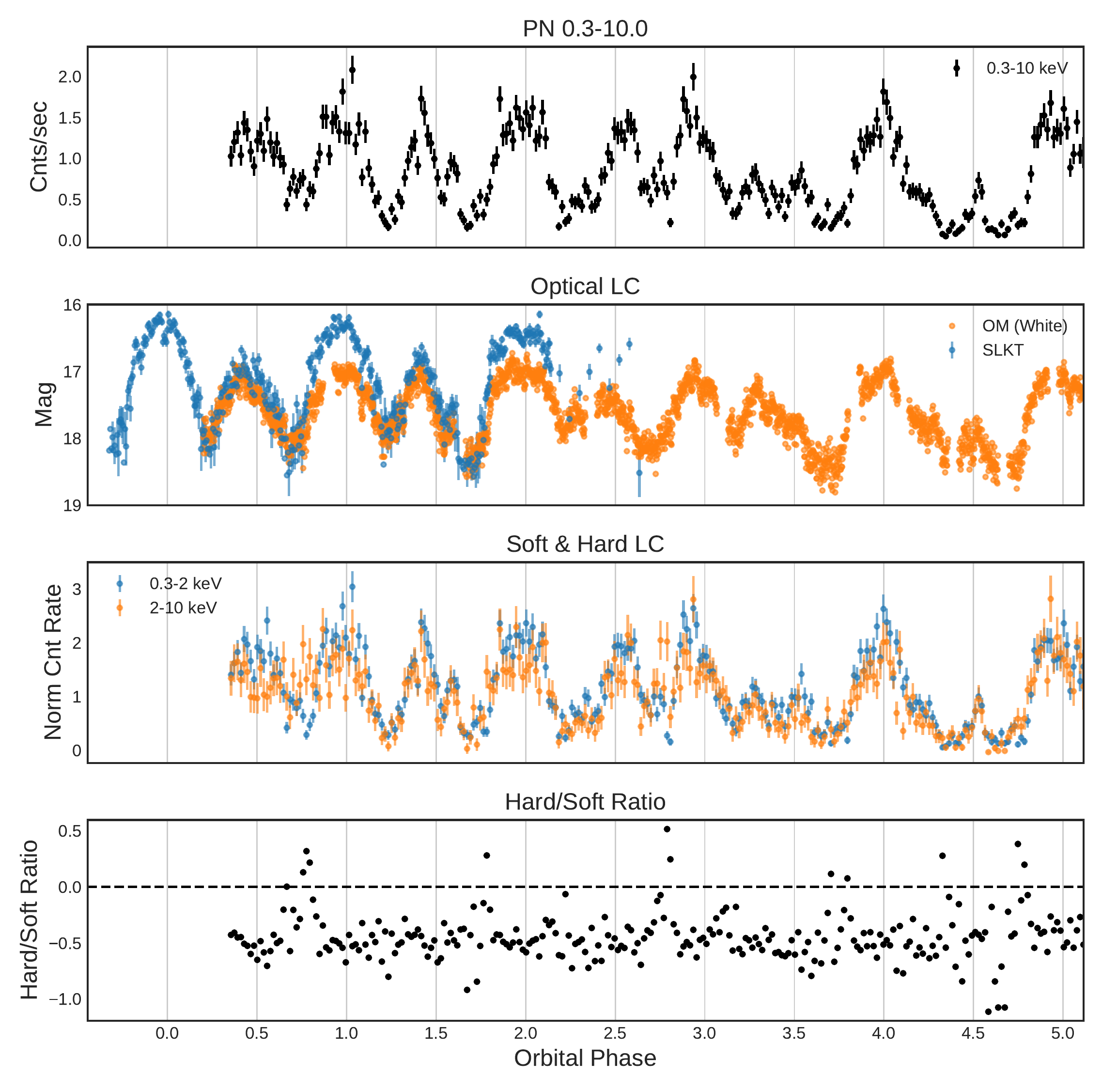}
    \caption{Light curves around the time of the \textit{XMM-Newton} observations. The top panel shows the full 0.3--10.0 keV light curve. The middle panel shows the optical light curves using data from the OM, the SLKT, and from members of the AAVSO community. The second panel shows the X-ray light split into 2 bands - soft (0.3--2 keV) and hard (2--10 keV). The bottom panel shows the ratio of these light curves, and highlights when we see a hard X-ray excess. Orbital phase has been calculated using the ephemeris described in the text.}
    \label{fig:full_lc}
\end{figure*}

\subsection{X-ray light curves}
Light curves were extracted for 3 energy ranges - the full energy range of the detectors (0.3-10.0 keV), a soft energy range of 0.3-2.0 keV, and a hard energy range of 2.0-10.0 keV. The Hardness ratio (($F_{2-10}-F_{0.3-2}$)/($F_{2-10}+F_{0.3-2}$); \citealt{2015A&A...583A.130W}) for the duration of the observations was also computed. The top panel of Figure~\ref{fig:full_lc} shows the 0.3-10.0 keV light curve of \jtto\ phased using the ephemeris from Section~\ref{sec:eph}. A total of 8 X-ray maxima were detected over 5 orbits of observations. The light curve over a single orbital period is composed of three features - a primary maximum at $\phi=0$ which corresponds to the optical maximum, a secondary maximum which occurs at $\phi=0.4$, and a rapid change in the Hardness ratio from -0.5 to +0.3 at $\phi=0.75$. The secondary maximum is only clearly detected for the first 3 orbital periods of data, after which its strength diminishes rapidly.

\subsection{X-ray spectra}
\subsubsection{Spectral Extraction}
\begin{figure*}
    \centering
    \includegraphics[width=\textwidth]{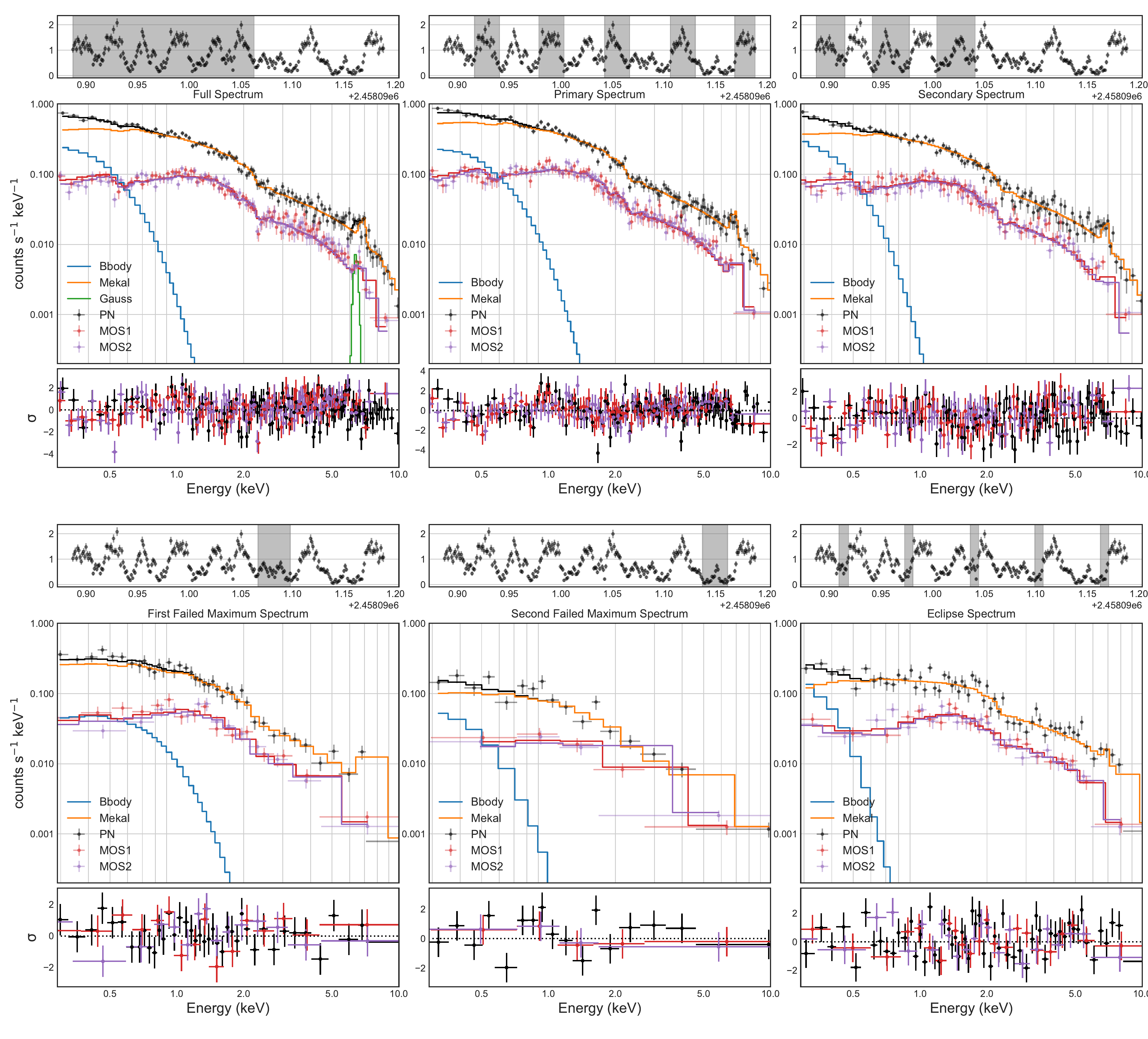}
    \caption{The extracted PN (black points), MOS1 (red points), and MOS2 (purple points) for 6 different time segments, along with the model residuals in units of $\sigma$. Above each spectrum is the 1-10 keV light curve from the PN instrument. The spectra in each panel were extracted using the highlighted time ranges. The best fit spectra to each epoch of data are also plotted in each panel as histograms, with the same colour scheme as the data. The components which are summed together to give the best fitting model to the PN instrument data are also shown in each panel, and consist of a blackbody component (blue), a diffuse hot plasma (\textsc{Mekal}; orange) and in one instance a 6.4 keV Gaussian emission component (green) }
    \label{fig:spectra}
\end{figure*}

Spectra covering the 0.3-10.0 keV were extracted for several different phase intervals as given by:
\begin{itemize}
    \item An X-ray spectrum constructed from all data up until the first missing secondary maximum (T (BJD)<245810.06). This is referred to as the ``half data'' set in the rest of the text.
    \item An X-ray spectrum of the primary maximum (data with $0.85<\phi<0.2$).
    \item An X-ray spectrum of the secondary maximum (data with $0.2<\phi<0.7$, but only for the first three orbital cycles).
    \item An X-ray spectrum of the first failed secondary maximum (data with $0.2<\phi<0.7$, but only for the fourth orbital cycle).
    \item An X-ray spectrum of the second failed secondary maximum (data with $0.2<\phi<0.7$, but only for the fifth orbital cycle).
    \item An X-ray spectrum of regions where the Hardness ratio was measured to be positive (data with $0.7<\phi<0.85$). This is referred to as the absorption dip spectrum from here onward.
\end{itemize}
In the case of each spectrum, the region for source extraction was chosen to be a circle with a radius of 24\arcsec around the target position. For the PN instrument, background spectra were extracted using a circular annulus centered on 13:21:32.46 +56:11:58.45 and with a radius of 57\arcsec. For both MOS instruments, background spectra came from a circular annulus centered on 13:21:33.25 +56:09:17.63 and with a radius of 96\arcsec. These spectra, along with the times of the X-ray observation they correspond to, are shown in Figure~\ref{fig:spectra}.

\subsubsection{Spectral Fitting}
The spectra were analysed using \texttt{Xspec} v12.10.1 \citep{xspec}. Each of the 6 extracted spectra were fit with a black body to account for the soft (<1.0 keV) component and a single temperature plasma emission model (\textsc{mekal} in \texttt{Xspec}; \citealt{1985A&AS...62..197M}; \citealt{1986A&AS...65..511M}; \citealt{1995ApJ...438L.115L}) to account for emission produced in the shock above the white dwarfs surface. Both components were absorbed by an interstellar absorber (\textsc{tbabs}, the Tuebingen-Boulder ISM absorption model; \citealt{tbabs}). Finally, the model was multiplied by a constant which was set to a value of 1 for the PN instrument, and allowed to vary for both the MOS1 and MOS2 instruments to allow for cross-instrument calibration. Such a model is a common starting place for describing the spectra of polars (e.g. \citealt{2005ApJ...620..422S}; \citealt{2015A&A...583A.130W}). We also included a Gaussian emission component at 6.4 keV to account for the common appearance of the Fe fluorescence feature at these energies in some accreting systems.

For the absorption dip spectrum, we added an additional partial covering absorption component (\textsc{pcfabs}), and froze all other parameters to their best fit values from modelling the spectrum of the primary maximum, under the assumption that the only difference between the absorption dip spectrum and the primary maximum spectrum should be additional absorption from the accretion stream.

The best fit parameters for these models were found using the default Levenberg-Marquardt algorithm in \texttt{Xspec} with a maximum number of 10000 evaluations allowed and a critical delta of $1\times10^{-4}$ required. The parameter space was then explored to obtain errors on the parameters by using the Goodman-Weare algorithm \citep{2010CAMCS...5...65G} for Markov Chain Monte Carlo's as implemented within \texttt{Xspec}. A total of 20 walkers were used, each of which were allowed to take 500,000 steps. The corner plots from the MCMC analysis of each of the spectra are included as an online dataset, while the corner plot from fitting the primary maximum is included in Appendix~\ref{sec:corner_plots}.

The results from fitting this model to each of the spectra are shown in Figure~\ref{fig:spectra}, and the best-fit parameter values are given in Table~\ref{tab:fitting_results}. We also include the unabsorbed, 0.3-10.0 keV X-ray luminosity (assuming a source distance of 760$\pm$30 pc) for just the plasma component of the model (that is, excluding the soft thermal emission from the white dwarf), which can be used as a stand-in for the mass accretion rate.

Of the six data sets which this model was applied to, only two have unacceptable an $\chi^{2}$ - the half data and primary maximum data. In the first instance, the poor $\chi^{2}$ can be attributed to the fact that the spectrum is the results of emission from both accreting magnetic poles of the WD, while the model is a single temperature plasma. As such, decomposing the spectrum into a primary and secondary spectrum improves this.

The cause of the $\chi^{2}$ of 305 for 275 degrees of freedom when fitting the spectrum of the primary maximum is more difficult to explain. In the above, we have assumed the hard X-ray emission comes from a single temperature plasma, the reality is likely more complex. The plasma should have a range of temperatures due to the ballistic stream coupling to the magnetic field across a range of angles, rather than at a single point. This is what likely leads to the high $\chi^2_{\rm R}$ value when modelling the primary maximum. As such, we have also fit the primary maximum data with the \textsc{mekal} replaced by \textsc{cemekl} \citep{1996ApJ...456..766S}, which allows for a multi-temperature plasma. The best fitting parameters and their errors (as estimated using the same methods as above) are given in Table~\ref{tab:fitting_results_cemekl}. The $\chi^{2}$ of 282 for 274 d.o.f is a significant improvement on the single temperature model, but there is a very strong anti-correlation between the index of the power-law emissivity function versus the maximum plasma temperature (as seen in Appendix~\ref{sec:corner_plots}), making it difficult to conclude anything physical from these models.

\begin{table*}
	\centering
	\caption{Model parameters from fitting each of the spectrum in Figure~\ref{fig:spectra} with an absorbed black body and plasma model. Errors are given at the $1\sigma$ level, and have been calculated as described in the text. Parameters marked with $a$ were frozen when fitting. The 0.3-10 keV X-ray luminosity of the \textsc{Mekal} component has been calculated assuming a source distance of 760$\pm$30 pc.}
	\label{tab:fitting_results}
	\begin{tabular}{lcccccc} % four columns, alignment for each
		\hline\hline
		Data Considered         & Half data & Primary Max & Secondary Max & Failed Max \# 1 & Failed Max \# 2 & Absorption dip \\
		\hline
		$n_{\rm H}$ ($\times10^{22}\:{\rm cm}^{-2}$) & <$0.005$ & <$0.008$ & <$0.01$ & <$0.06$ & <$0.15$ & 0.008$^{a}$\\
		$n_{\rm H, pcfabs}$ ($\times10^{22}\:{\rm cm}^{-2}$) & - & - & - & - & - & $2.1\pm0.4$\\
		CvrFract & - & - & - & - & - & $0.68\pm0.01$\\
		$kT_{\rm BB}$ (keV) & $0.078^{+0.01}_{-0.009}$ & $0.09\pm0.02$ & $0.06^{+0.01}_{-0.009}$ & <$0.25$ & <$0.27$ & 0.09$^{a}$\\
		$norm_{\rm BB}$ ($(\times10^{-6})$) &  $3.1^{+0.7}_{-0.5}$ & $2.8^{+0.7}_{-0.4}$ & $5^{+2}_{-1}$ & <$8$ & <$8$ & 2.8$^{a}$\\
		$kT_{\rm mekal}$ (keV) & $15\pm1$ & $13\pm1$ & $14\pm2$ & $7^{+2}_{-1}$ & $20^{+20}_{-10}$ & 13$^{a}$\\
		$norm_{\rm mekal}$ $(\times10^{-3})$ & $1.42\pm0.02$ & $1.69\pm0.03$ & $1.22\pm0.03$ & $0.74\pm0.04$ & $0.31\pm0.05$ & 1.69$^{a}$\\
		$C_{\rm MOS1}$ & $0.95\pm0.02$ & $0.94\pm0.02$ & $0.96\pm0.03$ & $0.96\pm0.07$ & $1.0\pm0.1$ & $0.92\pm0.05$\\
		$C_{\rm MOS2}$ & $0.96\pm0.02$ & $0.97\pm0.02$ & $0.94\pm0.03$ & $0.90\pm0.07$ & $1.0\pm0.1$ & $0.89\pm0.05$\\
		$L_{\rm MEKAL,0.3-10 keV}$ (erg/s) &  & $2.5\pm0.2\times10^{32}$& $1.8\pm0.1\times10^{32}$&  $1.1\pm0.1\times10^{32}$ & $0.42\pm0.05\times10^{32}$\\
		\hline
		$\chi^2$ (d.o.f) & 340 (294) & 305 (275) & 245 (230) & 52 (65) & 26.25 (19) & 185 (98)\\
		\hline\hline
	\end{tabular}
\end{table*}

\begin{table}
	\centering
	\caption{Results from applying the multi-temperature plasma model to the primary maximum data.}
	\label{tab:fitting_results_cemekl}
	\begin{tabular}{lc} % four columns, alignment for each
		\hline\hline
		Data Considered         Primary Max \\
		\hline
		$n_{\rm H}$ ($\times10^{22}\:{\rm cm}^{-2}$) & <$0.005$ \\
		$kT_{\rm BB}$ (keV) & $0.071^{+0.008}_{-0.009}$ \\
		$norm_{\rm BB}$ ($(\times10^{-6})$) &  $5.0^{+2.0}_{-1.0}$ \\
		$\alpha$ & $1.1^{+0.2}_{-0.1}$\\
		$kT_{max}$ (keV) & $41^{+13}_{-9}$\\
		$norm_{\rm mekal}$ $(\times10^{-3})$ & $4.2\pm0.5$\\
		$C_{\rm MOS1}$ & $0.94\pm0.02$\\
		$C_{\rm MOS2}$ & $0.97\pm0.02$\\
		\hline
		$\chi^2$ (d.o.f) & 282 (274)\\
		\hline\hline
	\end{tabular}
\end{table}

\subsection{X-ray absorption dip}
The spikes in the Hardness ratio occur at the same orbital phase during which absorption lines appear in the optical spectrum of \jtto. The X-ray spectra extracted during this orbital phase show that the spike in the hard-soft ratio are due to a significant decrease in the soft X-ray flux (as opposed to being an increase in the hard X-ray flux).

Modelling of this spectrum reveals that the decrease in the soft X-ray flux is due to a sudden increase in the absorption column between us and \jtto. The most likely cause of the sudden increase in absorption is the accretion stream passing through our line of sight, temporarily blocking the view of the accreting primary pole and absorbing a majority of the produced soft X-rays. Since modelling of the other extracted spectra allow us to put strong upper limits on the interstellar absorption column of $n_{\rm H}<0.01\times10^{22}$ cm$^{-2}$ in the direction of \jtto\, we can attribute the entirety of the measured value of $n_{\rm H}=(2.1\pm0.4)\times10^{22}$ cm$^{-2}$ to absorption by the accretion column. In terms of the system geometry, this absorption dip suggests the accretion stream is leading the companion star, as this soft X-ray absorption dip occurs before inferior conjunction of the companion.

If the single-to-noise of the individual absorption dip spectra were high enough, the measurement of the particle density of the column could be used to directly measure variations in the mass-accretion rate over the timescale of a single orbit. Unfortunately, these data do not have the sufficient S/N to do this, but it may be possible with future, more sensitive X-ray missions.

\subsection{Shock temperatures and magnetic field geometry}
The improvement of the multi-temperature plasma model over the single temperature model when modelling the primary maximum is in line with the connecting region between the ballistic stream and the primary poles magnetic field spanning a range of azimuth and radii. We are not able to derive strong constraints on the maximum plasma temperature, likely due to the low count rate at the highest energies of our spectrum. Further X-ray data taken at high energies (for example, with \textit{NuSTAR}) will be able to better constrain the highest shock temperature.

The very good agreement between the model and data for the secondary maximum suggests one of two things. Either the material which is feeding this pole comes from a very narrow connecting region, leading to a shock which is very close to uniform in temperature or, more likely, the spectrum does not have sufficient signal to differentiate between a single and multi-temperature plasma. Again, observations at a higher X-ray energy will help differentiate the two scenarios.

With the detection of two distinct maxima in the X-ray light curve, it is very likely that the white dwarf primary in \jtto\ is accreting onto both of its magnetic poles, as proposed by \cite{littlefield18}. If the magnetic field in \jtto\ were perfectly dipolar, one would naively assume that these maxima should be 180 degrees apart, or in other words, separated by 0.5 in orbital phase. This is very close to the observed phase separation of the two X-ray maxima in \jtto\ when both maxima are present and stable, suggesting the structure of WD's magnetic field might be reasonably approximated as dipolar. An additional test for this can be done by measuring the magnetic field of both poles. This is typically done by measuring the cyclotron harmonics in the optical spectrum of both accretion regions (as was done for e.g. V808 Aur; \citealt{2015A&A...583A.130W}). However, as highlighted by \cite{littlefield18}, measurement of the magnetic field in \jtto\ is complicated by significant smearing of the harmonics, hampering attempts to measure the magnetic field of both poles.

\subsection{Failed X-ray maximum}
The detection of 2 maxima per orbital phase during the first 3 orbits of \xmm\ data confirm the suggestion put forward by \cite{littlefield18} that accretion onto the WD in \jtto\ typically occurs via two distinct magnetic poles. Modelling of the observed X-ray spectra during these maxima reveal that both accretion columns have approximately the same temperatures in the shock in the accretion column, and that both polar caps of the WD are heated to the same degree.

The two failed secondary maxima in the optical light curve during the latter half of the \xmm\ observations coincide with a significant decrease in the amplitude of the secondary maxima in the X-ray light curve. Assuming the blackbody component of our models is coming from the WD surface, modelling of these two failed maxima show that the temperature of the WD surface was unchanged (to within 1$\sigma$) when compared with the derived temperature when the secondary maximum was present. On the other hand, the shock temperature exhibits a rapid decrease between the times when the secondary maximum is present ($kT=17^{+4}_{-2}$ keV) and when it is not present ($kT=7^{+2}_{-1}$ keV during the first failed maximum, and completely unconstrained during the second failed maximum).

This suggests that accretion onto the second, less preferential magnetic pole decreases significantly but does not cease entirely. The primary maxima before these failed secondary maxima are not significantly brighter than the primary maxima which occur when the secondary maximum is fully present. This rules out the case that more material gets channeled onto the primary magnetic pole during the failed secondary maxima as, if this were the case, we would expect the primary maxima to increase in strength.

Rather, the data suggests an overall decrease in the mass transfer rate in the system, which leads to less material making it to the secondary magnetic pole while maintaining the same amount of material reaching the primary maximum. The cause of this decrease in the mass transfer rate is unclear, but may be related to activity on the surface of the secondary star. Such a model has been invoked to explain the transient two-pole accretion seen in QS Tel \citep{1995A&A...293..764S, 1996MNRAS.280.1121R} and MT Dra \citep{2002A&A...392..505S}.

\section{Optical Photometry}

\subsection{Light curves}

\begin{figure}
    \centering
    \includegraphics[width=\columnwidth]{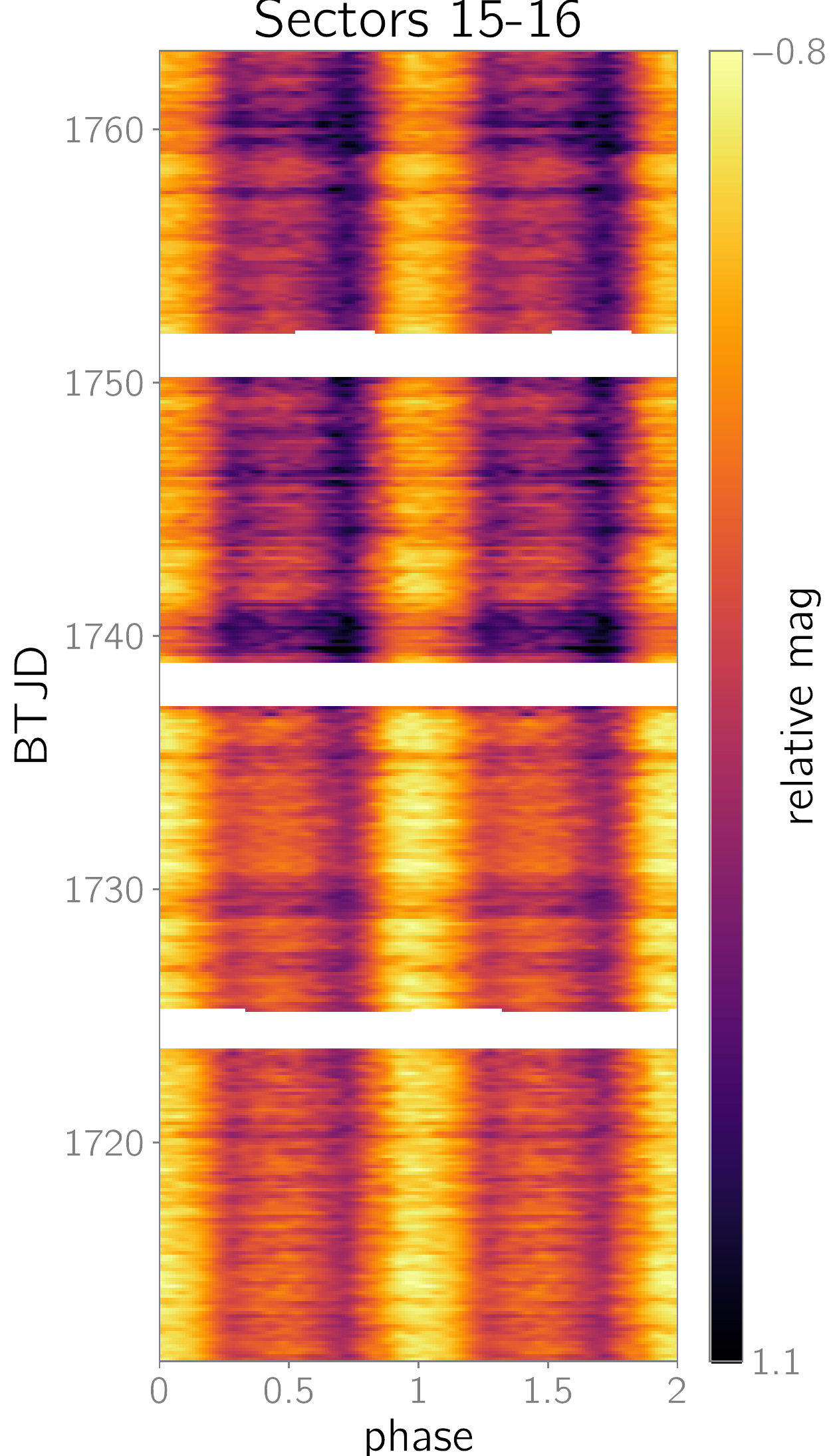}
    \caption{The full TESS light curve, phased to the orbital period. The width of the sliding window is one-eighth of a day. The three horizontal white bands indicate gaps due to data downlink. The central gap coincides with the transition from Sector 15 to Sector 16, and because of the changed spacecraft pointing, there is a brightness discontinuity at that gap.}
    \label{fig:tess_2D}
\end{figure}

The \tess\ light curve (which can be seen in Figure~\ref{fig:tess_2D}) is generally consistent with previous optical observations of the system \citep{littlefield15, littlefield18}, except that the secondary maximum does not stand out as prominently. It has a lower amplitude, and often blends with the primary maximum. This is probably attributable to differences in the cyclotron continua of the two poles; time-resolved spectroscopy of a binary orbit \citep{littlefield18} shows that the continuum for the primary pole shows more variability at the longer wavelengths which \tess\ is sensitive to. The variability of the second pole's cyclotron continuum increases at shorter wavelengths, thereby explaining why the secondary pole is more pronounced in optical observations than in the near-infrared TESS bandpass.

\begin{figure}
    \centering
    \includegraphics[width=\columnwidth]{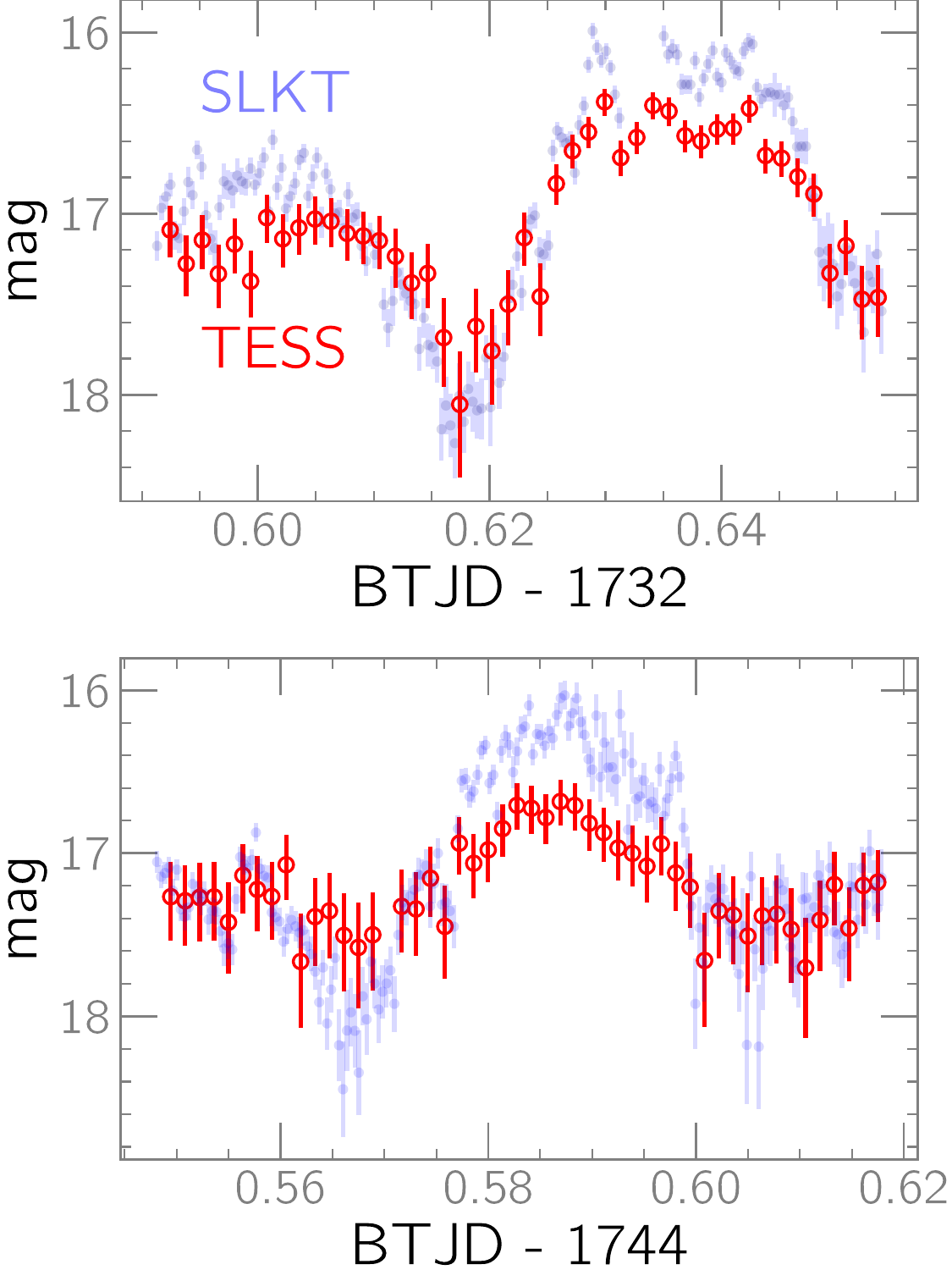}
    \caption{Comparison of simultaneous light curves of \jtto\ obtained with \tess\ and the SLKT. The SLKT data were obtained without a filter and use a Johnson $V$ zeropoint, and the \tess\ data were converted to an instrumental magnitude, with an arbitrary offset added. The \tess\ light curve shows a decreased amplitude of variability, attributable to a combination of blending and a bandpass difference.}
    \label{fig:SLKT}
\end{figure}

We obtained one ground-based light curve of \jtto\ with the SLKT during each sector of \textit{TESS} observations, with the aim of ascertaining whether the variability observed in the \tess\ bandpass is consistent with the variability observed in previous optical studies. The overall shape of the light curve is consistent across the two bandpasses as can be seen in Figure~\ref{fig:SLKT}, and the primary photometric maximum in the \tess\ light curve is the same as the primary photometric maximum in the optical light curve. However, the relative amplitude of the variation is reduced in the \tess\ bandpass, a likely consequence of blending with nearby sources. Additionally, the rapid flickering in the SLKT light curve is not always apparent in the \tess\ data, possibly because the time resolution of the SLKT was superior by a factor of $\sim$4.

%\subsection{Power spectrum}
%Lomb-Scargle power-spectral analysis of the TESS light curve reveals only the orbital frequency and its harmonics. We scrutinized the \tess\ light curve for evidence of quasi-periodic oscillations (QPOs) by subtracting a model of the orbital waveform and calculating a trailed power spectrum of the residuals; however, no QPOs were present.

\subsection{Optical Ephemeris}\label{sec:eph}

\begin{figure*}
    \centering
    \includegraphics[width=\textwidth]{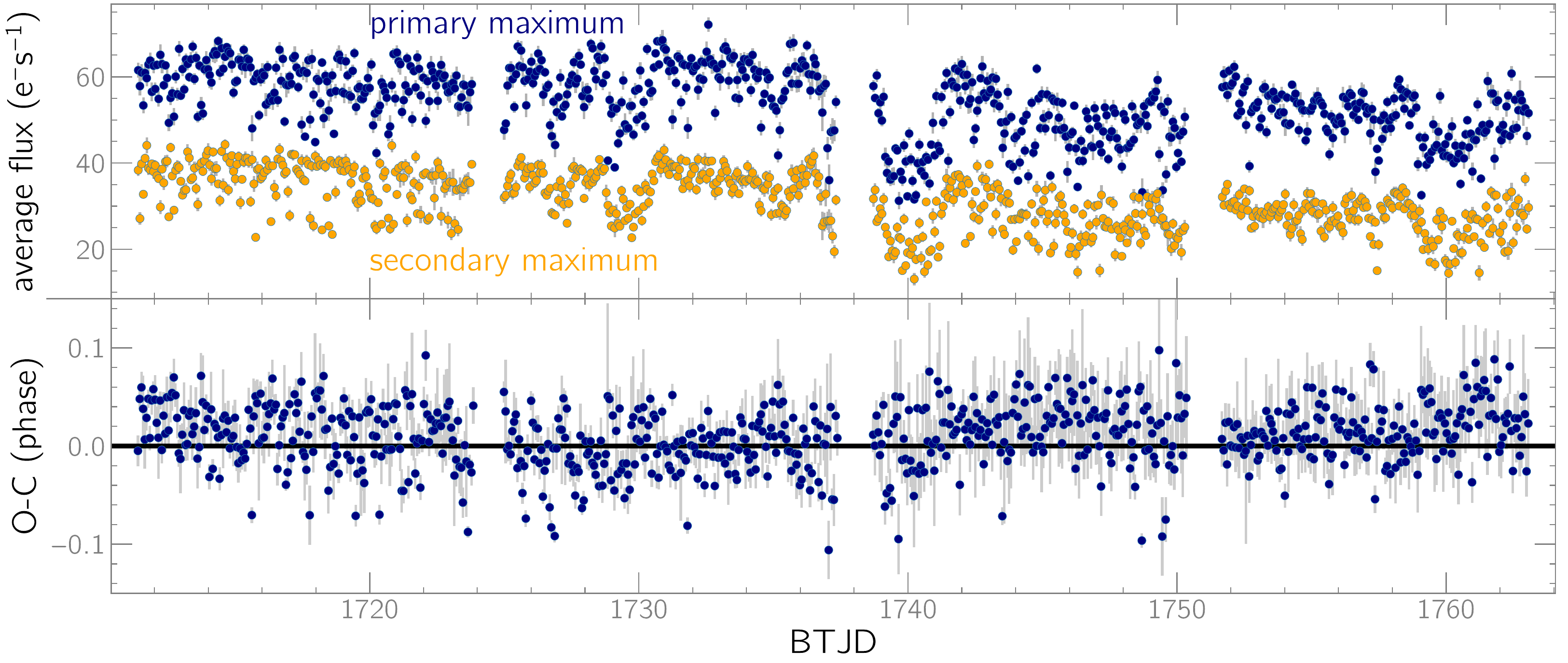}
    \caption{{\bf Top:} The brightness of each accretion region during Sectors 15 and 16 of \tess. For both the primary and secondary maxima, we calculated the average flux within $\pm$0.1 phase units of the expected phase of maximum light. The secondary maxima show particularly erratic variability. {\bf Bottom:} O$-$C of the primary photometric maxima with respect to the orbital period from \citet{littlefield18} and a reference time $T_{0}$ from the \tess\ dataset. The O$-$C values display a slow, apparently aperiodic drift that does not correlate with the brightness of either the primary or secondary maximum.}
    \label{fig:TESS_maxima}
\end{figure*}

A common method of measuring the orbital period in a polar is to measure the recurrence interval of a well-defined feature in the light curve. At first glance, the primary photometric maximum of V496~UMa is ideal for this purpose. We fit third-order polynomials to each of the primary photometric maxima, visually inspected the resulting fits to ensure their adequacy, and used each polynomial to calculate the time of maximum flux for each peak. We used a Monte Carlo procedure to estimate the uncertainty of each timing and calculated a best-fit linear ephemeris of \begin{equation}
    T_{max}[BJD] = 2458722.01138(4) + 0.0632329(2)
    \label{eq:ephem}
\end{equation} for the \textit{TESS} data. This period is very different from the period of 0.063235199(40)~d reported in \citet{littlefield18}. Inspection of the residuals from the ephemeris (Fig.~\ref{fig:TESS_maxima}) suggest a possible explanation for this discrepancy. The residuals from Eq.~\ref{eq:ephem} show a systematic curvature consistent with a gradual, aperiodic phase shift of the primary photometric maximum. On relatively short timescales ($\lesssim2$ weeks), a linear ephemeris can compensate for this phase drift with a change in the apparent orbital period. For example, the residuals in Fig.~\ref{fig:TESS_maxima} are clustered into four groups (each corresponding to one spacecraft orbit), and the best-fit periods for each of the four groups differed from the \citet{littlefield18} period by up to $\sim\pm1$~s.

It is obviously unphysical for the binary orbit to change by such a large amount in such a short time, but it is possible for the position of the cyclotron-emitting region to drift across the face of the WD. Such behavior is expected in a polar, as the location of the accretion region is not fixed to the binary frame and depends on which field lines are channeling the infalling matter. Variations in the mass-transfer rate could therefore cause a phase shift of the accretion region, in which case one would also expect such a change to produce observable luminosity variations. However, the O$-$C does not show any significant correlation with the system's brightness. 

The detection of this oscillation is reminiscent of the aperiodic drift in the optical maxima of the intermediate polar FO~Aqr, identified by \citet{kennedy17} using \textit{Kepler K2} data. \citet{kennedy17} demonstrated how this effect can frustrate attempts to precisely measure the orbital period, but it has not been previously reported in a synchronous polar. The drift observed in \jtto\ is sufficiently small and gradual that poorly sampled ground-based observations might struggle to distinguish between this effect and an inaccurate measurement of the orbital period. It is unclear whether this phase drift is a persistent feature of \jtto\ or whether it occurs in polars generally, and but as \tess\ continues to observe polars, it will be possible to search for this effect in other systems.

\section{Discussion}

The failed secondary maxima in the \tess\ observations can be broadly classified into two categories: those that correlate with the system's luminosity, and those that do not. \citet{littlefield18} noted that at optical wavelengths, the primary photometric maximum appeared to be unaffected by nearby failed secondary maxima, and a number of the failed maxima in the \tess\ light curve share this property. However, the \tess\ light curve shows multi-day-long depressions near BTJD = 1730 and BTJD = 1740, and during these episodes of reduced mass transfer, the failed secondary maxima are much more frequent, occurring in a majority of the orbital cycles.

Although the failed maxima in \citet{littlefield18} created the impression that there is a relatively clear dichotomy between normal and failed maxima, the extensive \tess\ data demonstrate that this is not so in the near-infrared \tess\ bandpass. On the contrary, the secondary maxima observed by \tess\ show such a wide range of behaviours that it can be difficult to categorize some of the maxima. Contamination from a nearby background star of similar brightness further complicates matters, since it means that if \jtto\ were to become undetectably faint, there would still be a weak signal at its position.

\jtto\ joins a growing list of polars which display dips due to obscuration of the soft X-ray producing region by the ballistic stream. The \textit{XMM-Newton} observations strongly suggest that it is a two-pole accretor with highly intermittent accretion onto its secondary magnetic pole, and that this intermittent behaviour is being driven in changes in the mass transfer rate from the donor star. The \tess\ light curve does not show the missing secondary maxima as clearly as previous optical observations, and also suggest that the ephemeris derived from short intervals of observations may be unreliable. We speculate that this is because the cyclotron spectrum of the secondary accretion region from \citet{littlefield18} is quite blue, so its relative contribution in the \tess\ bandpass is relatively low.

\subsection{Comparison with other systems}
Two-pole accretion within polars is an informative phenomenon, as it allows us to study both multiple accretion regions on the white dwarf's surface, and probe a larger volume of the WDs magnetosphere over the single pole case. This is particularly powerful in systems with transient two pole accretion, where the location within the magnetic field which the accretion stream is probing varies. Two-pole accretion is not uncommon, and there are numerous instances in the literature of polars that have switched between one- and two-pole accretion, the prototype polar AM Her \citep{1985A&A...148L..14H}, MT Dra \citep{2002A&A...392..505S}, and QS~Tel \citep{1996MNRAS.280.1121R} being excellent examples. It is not always clear why the number of active poles changes, though \citet{1996MNRAS.280.1121R} and \citet{2002A&A...392..505S} considered two hypotheses for QS~Tel and MT Dra, respectively: a change in the mass-transfer rate and asynchronous rotation. In the former, the accretion stream's ram pressure depends on the mass-transfer rate, causing the stream to travel deeper into the magnetosphere at higher mass-transfer rates. In the latter, the accretion stream would latch onto different magnetic field lines due to the differential rotation of the magnetosphere; these variations would occur at the beat frequency between the binary orbital frequency and the WD's spin frequency, which ranges from a few days to $\sim$2 months in the known asynchronous polars. 

The data presented here confirm that the transient two-pole accretion in \jtto\ is not due to asynchronous rotation of the WD. If this were the case, we would have expected correlated changes in the primary and secondary maxima of the X-ray light curve, and a long-term periodicity in the \textit{TESS} light curve \citep[as observed in CD~Ind;][]{hakala19, littlefield19, 2020AdSpR..66.1123M}. The absence of these effects suggest the driving force between the variability in the secondary pole may be more akin to what is occurring in AM Her and other synchronous polars. 

Although the variable mass-transfer-rate explanation is more promising, it is has its own shortcomings. If we use \jtto's time-averaged optical brightness as a proxy for its accretion rate, then there is no consistent relation between its overall accretion rate and the failed secondary maxima; in Figs.~\ref{fig:tess_2D}~and~\ref{fig:TESS_maxima}, the failed maxima are common during a dip near BTJD=1740, but they also occur sporadically when \jtto\ is brightest. The lack of such a correlation is reminiscent of the behaviour of AM~Her, whose accretion geometry does not always correlate strongly with the mass-transfer rate \citep{2020A&A...642A.134S}. Similarly, \citet{HY_Eri} found that HY~Eri remains in a two-pole-accretion state even when the mass-transfer rate varies by three orders of magnitude. However, BL~Hyi provides a countervailing example, as it undergoes two-pole accretion at enhanced accretion rates but one-pole accretion in its low states \citep{1989A&A...223..179B}.

A related scenario considered by \citet{1996MNRAS.280.1121R} to explain why QS Tel changed between one- and two-pole accretion was that the accretion stream can be fragmented into discrete blobs of varying densities. In such a case, the lifetime of any individual blob depends on its density, with the densest blobs surviving longer and traveling deeper into the WD's magnetosphere. Based on this picture, \citet{1996MNRAS.280.1121R} proposed that low-density material becomes magnetically entrained shortly after it leaves the donor star, producing a hard X-ray-emitting region, while higher-density blobs travel to a secondary accretion region with a softer spectrum. In this scenario, a temporary reduction in the number of dense blobs could interrupt accretion onto the second pole. It is worth noting that AM Her and MT Dra do conform to the \citet{1996MNRAS.280.1121R} scenario; when their secondary poles are active, they have softer spectra than their primary poles \citep{2020A&A...642A.134S, 2002A&A...392..505S}.

However, if this mechanism were at play in \jtto\, we would have expected to observe a pronounced difference in the X-ray hardness of the two poles. At most, \jtto\ may display a slight enhancement of soft X-ray emission coming from the secondary pole when is active and accreting; the blackbody component may be slightly higher for the secondary maximum than for the primary maxiumum, as given in Table~\ref{tab:fitting_results}. However, the enhancement is in no way definitive, and higher signal-to-noise spectra are required to tell. Indeed, the differences between the secondary and primary spectra are not nearly as extreme as in the case of AM Her, suggesting \jtto\ is not undergoing blobby accretion.

While \jtto\ does not offer an obvious answer as to why the secondary maximum occasionally disappears, this system stands out because of the time scale over which this takes place. For some of the best studied polars which undergo mode switching, the accretion geometry appears to be relatively stable during individual observing epochs. In contrast, the geometry in \jtto\ changes over the course of a single orbit, as shown here, and few synchronous polars have been observed to show such rapid optical changes in the accretion rate onto a secondary pole. One such example is DP~Leo \citep{DP_Leo}, which showed an intermittent secondary photometric maximum in optical photometry. However, there is insufficient data about this phenomenon in DP~Leo to draw any robust comparisons with \jtto. 

Whichever mechanism is altering the accretion geometry in \jtto\ (and presumably DP~Leo) varies over a short timescale. The most obvious culprit are stellar spots on the secondary star moving across the L1 point. Verification that this is driving a change in the mass transfer rate from the companion star would require optical spectroscopic and photometric observations in which the companion star dominates. This is a difficult task when accretion structures and cyclotron harmonics are present in the system, and \citet{littlefield18} were unable to detect the secondary star spectroscopically when \jtto\ was in a high state. As such, \jtto\ should be monitored frequenctly for the onset of a low state, at which point studies of the secondary, and indeed measurement of the WD's magnetic field through Zeeman splitting of the absorption lines created in the WD photosphere, would become possible.

\section*{Acknowledgements}

We thank the \tess\ mission-operations personnel, particularly George Ricker and Roland Vanderspek, for scheduling DDT observations of \jtto\ after a last-minute pointing change made it unexpectedly observable during Sectors 15 and 16. We thank the Krizmanich Family for their generous donation to the
University of Notre Dame that funded the Sarah L. Krizmanich Telescope. 

M.R.K. acknowledges support from the ERC under the European Union's Horizon 2020 research and innovation programme (grant agreement No. 715051; Spiders), the Royal Society in the form of a Newton International Fellowship (NIF No. NF171019), and the Irish Research Council in the form of a Government of Ireland Postdoctoral Fellowship (GOIPD/2021/670: Invisible Monsters).

This work made use of \textsc{Astropy} \citep{astropy:2013,astropy:2018}, \textsc{Corner} \citep{corner}, and \textsc{Pyxspec}.

\section*{Data availability}
The raw X-ray data are available through the \xmm\ Science Archive, while the TESS data are available through the Barbara A. Mikulski Archive for Space Telescopes. The exact X-ray spectra and light curves used in this paper can be found at the following permanent repository: \url{https://zenodo.org/record/5746735}.

%%%%%%%%%%%%%%%%%%%%%%%%%%%%%%%%%%%%%%%%%%%%%%%%%%



%\bibliographystyle{mnras}

\bibliographystyle{mnras}
\bibliography{j1321.bib} % if your bibtex file is called example.bib

%%%%%%%%%%%%%%%%%%%%%%%%%%%%%%%%%%%%%%%%%%%%%%%%%%

%%%%%%%%%%%%%%%%% APPENDICES %%%%%%%%%%%%%%%%%%%%%

\appendix

\section{Corner Plots}\label{sec:corner_plots}

\begin{figure*}
    \centering
    \includegraphics[width=\textwidth]{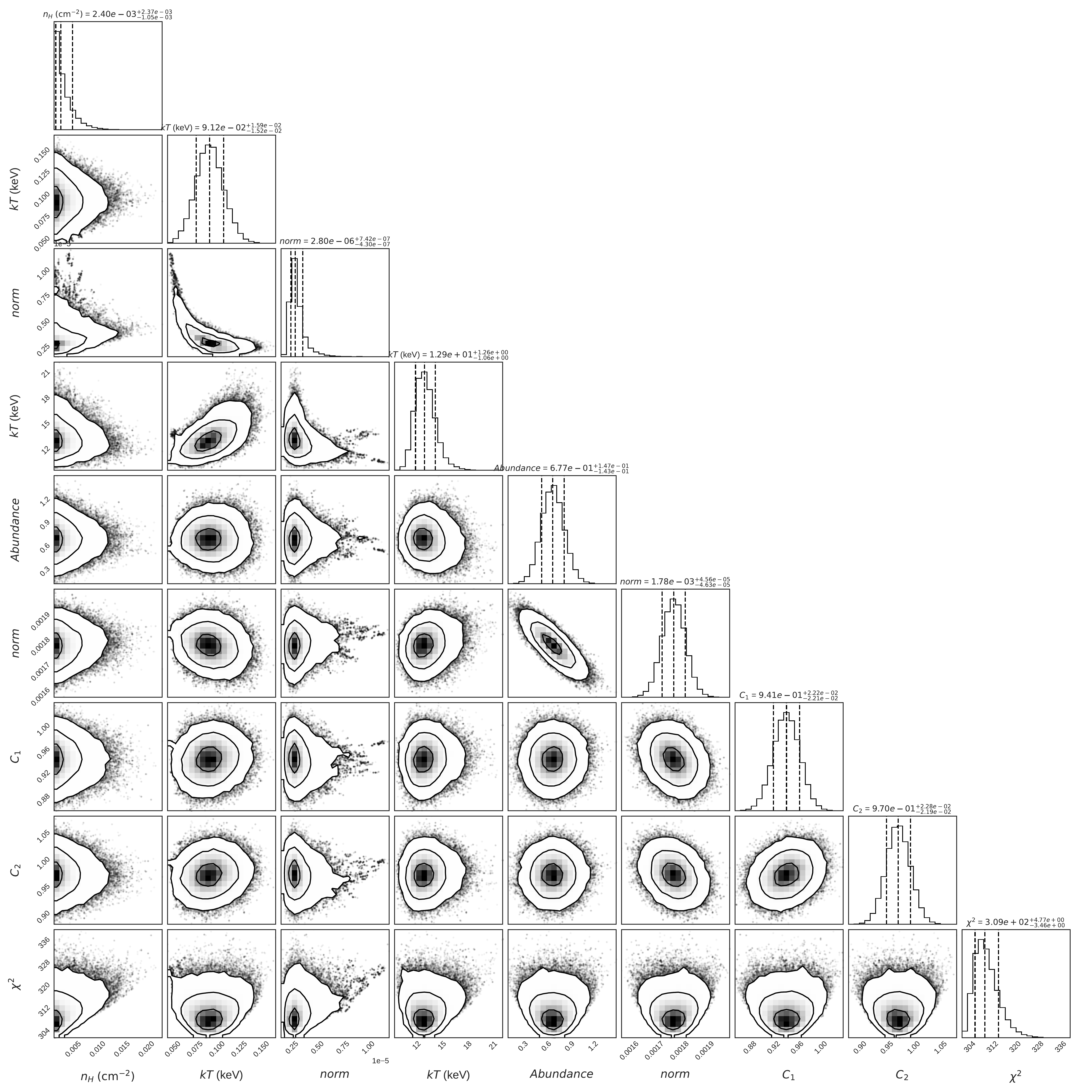}
    \caption{Corner plot from the MCMC analysis of the primary maximum spectrum using the single temperature plasma and thermal black body model.}
\end{figure*}

\begin{figure*}
    \centering
    \includegraphics[width=\textwidth]{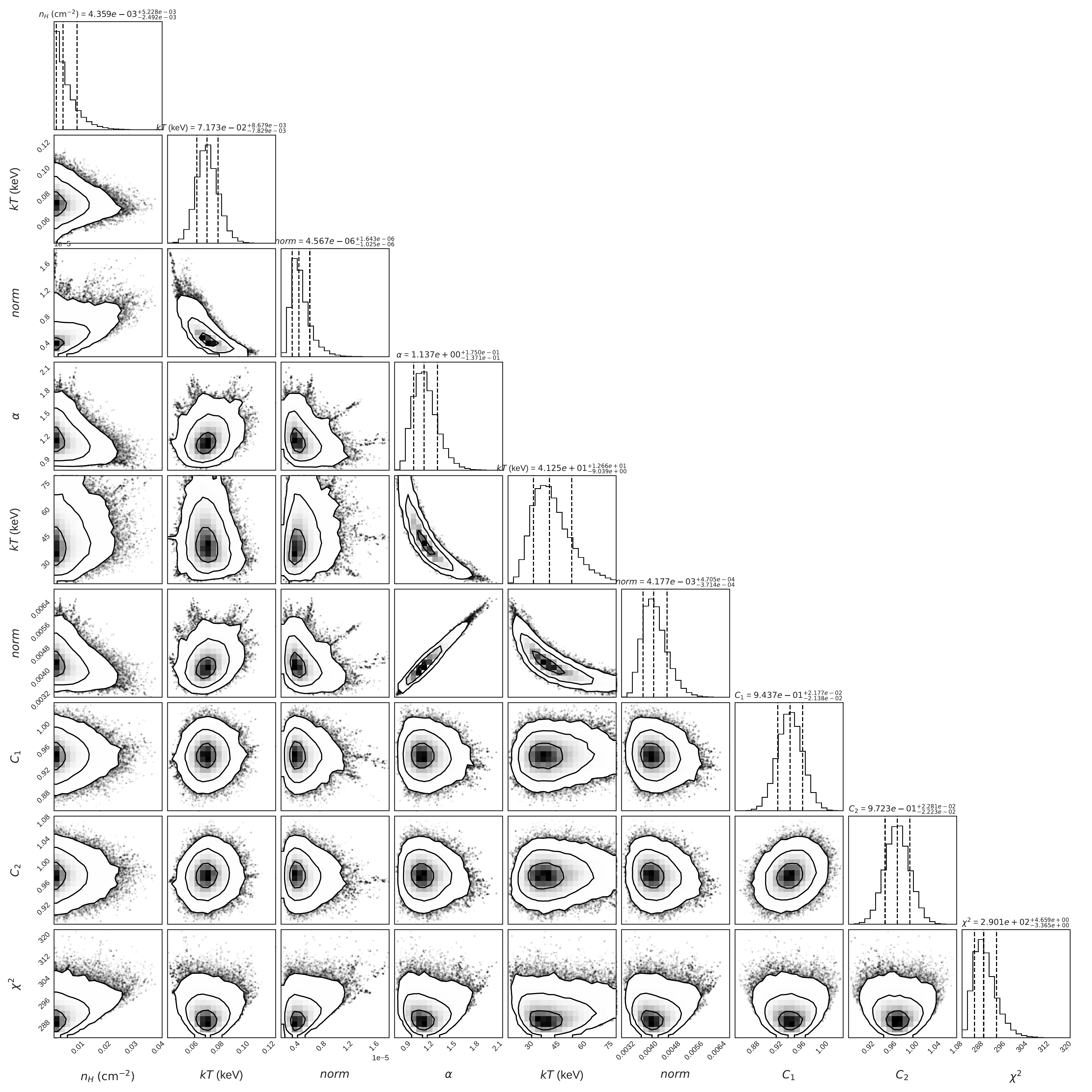}
    \caption{Corner plot from the MCMC analysis of the primary maximum spectrum using the multi temperature plasma and thermal black body model.}
\end{figure*}

%%%%%%%%%%%%%%%%%%%%%%%%%%%%%%%%%%%%%%%%%%%%%%%%%%

% Don't change these lines
\bsp	% typesetting comment
\label{lastpage}
\end{document}